\documentclass[preprintnumber,aps,nofootinbib,showpacs,showkeys]{revtex4-2}%WI.bib

\usepackage{array}
\usepackage{booktabs}
\usepackage{tabu}
\usepackage{dcolumn}
\usepackage{amsmath}
\usepackage{amsfonts}
\usepackage{amssymb}
\usepackage{graphicx}
\usepackage{subfigure}
\usepackage{graphicx}% Include figure files
\usepackage{dcolumn}% Align table columns on decimal point
\usepackage{bm}% bold math
\usepackage{xcolor}

\begin{document}

\title{Exploring new subclass of k-inflation: tachyon inflation in $R+\eta T$ gravity model}
\author{Abolhassan Mohammadi}
   \email{a.mohammadi@uok.ac.ir; abolhassanm@gmail.com}
     \author{Fardin Kheirandish}
     %\email{fkheirandish@uok.ac.ir}

   \affiliation{Department of Physics, Faculty of Science, University of Kurdistan, Pasdaran Street, P.O. Box 66177, Sanandaj, Iran.}
%    \affiliation{Physics}
%\author{Mr Physics}
%   \email{ggg@gmail.com}
%   \affiliation{Department}

\begin{abstract}
%Considering the alternative gravity theory $f(R,T)$, where there is a non-minimal coupling between the matter and curvature, the scenario of tachyon inflation is studied. After imposing the slow-roll approximations, the model is investigated in detail for different types of the potential as the power-law, the generalized T-mode, and the inverse coshyperbolic. Applying observational data on Python coding, a range for the free parameters of the model are determined for which any choice in the obtained range, the model perfectly meets the data.

It is explained that any scalar field in $f(R,T)$ gravity model could present a new subclass of the k-essence model for inflation. While the case of the quintessence has been studied, there is an empty gap required to be filled by investigating more types of field models. Here, we are aiming to partially fill the gap and concentrate on the tachyon field. In this study, we explore the phenomenon of tachyon inflation in the context of the alternative gravity theory $f(R,T) = R + \eta T$. By applying slow-roll approximations, we examine the model in detail for three different potentials: power-law, generalized T-mode, and inverse hyperbolic cosh. Using observational data and Python coding, we determine a range of values for the model's free parameters that allow it to fit the data perfectly. In essence, this study provides a comprehensive analysis of tachyon inflation in the $R + \eta T$ gravity theory, offering new insights into this alternative framework and its potential to explain tachyon inflation.
\end{abstract}

\date{\today}
\maketitle
%\tableofcontents

%%%%%%%%%%%%%%%%%%%%%%%%%%%%%%%%%%%%%%%%%%
%%%%%%%%%%%%%%%%%%%%%%%%%%%%%%%%%%%%%%%%%%
%%%%%%%%%%%%%%%%%%%%%%%%%%%%%%%%%%%%%%%%%%
%%%%%%%%%%%%%%%%%%%%%%%%%%%%%%%%%%%%%%%%%%
%%%%%%%%%%%%%%%%%%%%%%%%%%%%%%%%%%%%%%%%%%
%%%%%%%%%%%%%%%%%%%%%%%%%%%%%%%%%%%%%%%%%%
%%%%%%%%%%%%%%%%%%%%%%%%%%%%%%%%%%%%%%%%%%
%%%%%%%%%%%%%%%%%%%%%%%%%%%%%%%%%%%%%%%%%%
%%%%%%%%%%%%%%%%%%%%%%%%%%%%%%%%%%%%%%%%%%
\section{Introduction}\label{Sec_intro}
Einstein's general theory of relativity, proposed in 1915, is the most well-known and accepted theory of gravity. The theory explains how gravity affects objects and how light bends when it passes near a massive object. Einstein's theory of general relativity has been tested numerous times to see if it works. Aside from Einstein's theory of gravity, there have been serious attempts to introduce alternative theories of gravity \cite{Golanbari:2014cva,Golanbari:2014wxa,Golanbari:2014zua,Aghamohammadi:2013eja,Saaidi_2012,Saaidi:2012ri,Saaidi:2012qp,Saaidi:2011zza,Saaidi:2010ju}. The main reason for all of this effort is that general relativity cannot fit cosmological data without including dark energy and dark matter. The $f(R,T)$ theory of gravity is one of these alternative theories on which we will concentrate. This theory, in general, a non-minimal coupling between the matter sector and the curvature could exist by including mixing terms like $R T$. Harko et al. \cite{Harko:2011kv} first proposed the theory, which has since been used to study various astrophysical and cosmological topics \cite{Rosa:2022osy,Goncalves:2022ggq,Goncalves:2021vci,Rosa:2021teg,Bhatti:2016caw,Zaregonbadi:2016xna,Moraes:2019pao,Alves:2016iks}; including inflation \cite{Bhattacharjee:2020jsf,Gamonal:2020itt,Taghavi:2023ptn,Ossoulian:2023moq}. \\

Following the initial proposal of an inflationary scenario \cite{starobinsky1980new,Guth:1980zm,albrecht1982cosmology,linde1982new,linde1983chaotic}, the scenario has been modified and developed in various ways and within various gravity theories \cite{Barenboim:2007ii,Franche:2010yj,Unnikrishnan:2012zu,Rezazadeh:2014fwa,Saaidi:2015kaa,
Fairbairn:2002yp,Mukohyama:2002cn,Feinstein:2002aj,Padmanabhan:2002cp,Aghamohammadi:2014aca,
Spalinski:2007dv,Bessada:2009pe,Weller:2011ey,Nazavari:2016yaa,
maeda2013stability,abolhasani2014primordial,alexander2015dynamics,Tirandari:2018xyz,
maartens2000chaotic,golanbari2014brane,Mohammadi:2020ake,Mohammadi:2020ctd,Mohammadi:2015upa,Mohammadi:2015jka,
berera1995warm,berera2000warm,hall2004scalar,Sayar:2017pam,Akhtari:2017mxc,Sheikhahmadi:2019gzs,Rasheed:2020syk,
Mohammadi:2018oku,Mohammadi:2019dpu,Mohammadi:2018zkf,Mohammadi:2019qeu,Mohammadi:2020ftb,
Mohammadi:2021wde,Mohammadi:2021gvf,Mohammadi:2022vru,Mohammadi:2022fiv}. According to a prevalent theory, a scalar field—the dominant element of the cosmos at the time—is what propels inflation. It is raised above the potential and slowly descends toward the potential's lowest point. The potential does not significantly change throughout this slowly rolling, and a quasi-de Sitter expansion is created \cite{Linde:2000kn,Linde:2005ht,Linde:2005vy,Linde:2004kg,Riotto:2002yw,Baumann:2009ds,Weinberg:2008zzc,Lyth:2009zz,Liddle:2000cg}. There are various alternatives for the scalar field known as inflaton, which controls inflation, including the canonical scalar field, the tachyon field, and the DBI field.
The data \cite{Planck:2013jfk,Ade:2015lrj,Akrami:2018odb} have given the inflation scenario a ton of support and made it the core of all cosmological models. \\

The introduced field models above are taken as subclasses of the k-essence model. The k-essence model was introduced as an alternative to the primes models of dark energy, i.e. cosmological constant and quintessence model, which both are faced two major challenges: fine-tuning and cosmological coincidence problem. Seeking to solve the problem led to the introduction of another dynamical dark energy with a Lagrangian of the form $p(\phi, X)$, where $X = - \nabla_\mu \phi \nabla^\mu \phi / 2$. This type of field is inspired by the effective field theories in string theory, and was first proposed in \cite{Armendariz-Picon:2000ulo} for explaining the late-time acceleration phase of the universe, following its earlier successful application to explain the early acceleration phase with the name of k-inflation \cite{Garriga:1999vw}. It has been found that the model demonstrates a generic way to solve the cosmological coincidence problem and also there is no requirement for the fine-tuning of initial conditions \cite{Armendariz-Picon:2000ulo}. The k-essence is actually a broad class in which the quintessence is a special case. Other well-known subclasses are the tachyon field and DBI models, both of which have been studied as promising models of cosmic inflation. \\

In our model, besides the matter field Lagrangian, $\mathcal{L}$, the trace of the energy-momentum tensor, $T$, is also included in the action. Because the matter field is usually assumed to have $\mathcal{L} = p$ and $T = \rho - 3p$, the energy density and pressure are combined in the Lagrangian. Therefore, as one takes the energy density and the pressure as the ones of quintessence, tachyon, or DBI, there would be new combinations of $X$ and $\phi$. In another word, there would be a new subclass of the k-essence model with new Lagrangian $p(\phi,X)$, worthy to be investigated in more detail. For instance, taking the tachyon field, the new Lagrangian is given by
\begin{equation}\nonumber
p(\phi, X) = - V(\phi) \; \left[ \; {\lambda \over 2 \sqrt{1 - 2X}} 
                               + \Big( {3 \over 2} \lambda + 1 \Big) \; \sqrt{1 - 2X} \; \right].
\end{equation}
While the consistency and validity of the quintessence model in $R + \eta T$ gravity model to explain cosmic inflation has been considered, there is no investigation on other field models such as the tachyon field (refer to \cite{Choudhury:2015hvr} for a review on tachyon field). Tachyon inflation is a type of inflationary cosmology that has gained attention as a possible explanation for the early universe's rapid expansion. Tachyon inflation in Einstein's gravity has been studied extensively, and the result showed also a great consistency with the data. This success of tachyon inflation and also the temptation of investigating a new subclass of k-essence motivate us to investigate the tachyon field in $R + \eta T$ gravity model as a possible candidate to describe inflation. \\

%The tachyon field in the standard gravity model has been determined to be a valuable candidate for describing inflation. Its result has proven to be in good consistency with the data as well. This success of tachyon inflation and also the tempting of investigating a new subclass of k-essence motivate us to investigate the tachyon field in $R + \eta T$ gravity model as a possible candidate to describe inflation.

%Most of the work on inflation in $f(R,T)$ gravity theory concentrated on applying a canonical scalar field for the role of inflaton, and there are no attempts to consider other candidates such as tachyon field (refer to \cite{Choudhury:2015hvr} for a review on tachyon field).  In $f(R, T)$ gravity theory, the action for the gravitational field is modified to include a non-minimal coupling between curvature and matter, represented by the $"T"$ term. This results in a different evolution of the universe compared to standard inflationary models. In the current paper, we are going to consider the scenario of slow-roll tachyon inflation in the frame of $f(R,T)$ gravity theory. During this specific time, the potential energy density of the field plays the main role, and we consider the model for different types of the potential.      \\

The paper is organized as follows: there is a general and brief look at a specific case of the theory of $f(R,T)$ gravity in Sec.II. Next, the tachyon field is introduced in Sec.III, which plays the role of inflaton. The field energy density and pressure are inserted in the Friedmann equation, and they will be simplified by applying the slow-roll approximations. The perturbation parameters are introduced, which play an important role in comparing the model with the data. Then, in Sec.IV, the model is considered in detail for three types of potential and the free parameters of the model are determined by performing coding programming. Finally, the results are summarized in Sec.V.

%%%%%%%%%%%%%%%%%%%%%%%%%%%%%%%%%%%%%%%%%%
%%%%%%%%%%%%%%%%%%%%%%%%%%%%%%%%%%%%%%%%%%
%%%%%%%%%%%%%%%%%%%%%%%%%%%%%%%%%%%%%%%%%%
%%%%%%%%%%%%%%%%%%%%%%%%%%%%%%%%%%%%%%%%%%
%%%%%%%%%%%%%%%%%%%%%%%%%%%%%%%%%%%%%%%%%%
%%%%%%%%%%%%%%%%%%%%%%%%%%%%%%%%%%%%%%%%%%
%%%%%%%%%%%%%%%%%%%%%%%%%%%%%%%%%%%%%%%%%%
%%%%%%%%%%%%%%%%%%%%%%%%%%%%%%%%%%%%%%%%%%
%%%%%%%%%%%%%%%%%%%%%%%%%%%%%%%%%%%%%%%%%%
\section{Basic equations}\label{fRT_gravity}
In a general shape, the action of the $f(R,T)$ theory of gravity is assumed to be given as follows \cite{Harko:2011kv}
\begin{equation}\label{action}
S=\frac{1}{2 \kappa^2} \int f\left(R,T\right)\sqrt{-g}\;d^{4}x+\int{L_\mathrm{m}\sqrt{-g}\;d^{4}x}\, ,
\end{equation}
in which $R$ is known as the curvature scalar and $T$ stands for the trace of the energy-momentum tensor. $f(R,T)$ is an arbitrary function of curvature scalar $R$ and the trace $T$. The determinant of the spacetime metric $g_{\mu\nu}$ is indicated by $g$. The Lagrangian density for the matter sector is described by $\mathcal{L}_m$. The gravitational constant is given through the quantity $\kappa$ as $\kappa^2 = 8 \pi G$. \\
Taking variation of the action with respect to the metric leads to the field equation of the model, read as
\begin{equation}\label{fieldequation}
f_{R}\left( R,T\right) R_{\mu \nu } - \frac{1}{2}
f\left( R,T\right)  g_{\mu \nu }
+\left( g_{\mu \nu }\square -\nabla_{\mu }\nabla _{\nu }\right)
f_{R}\left( R,T\right) = \kappa^2 T_{\mu \nu}-f_{T}\left( R,T\right)
T_{\mu \nu }-f_T\left( R,T\right)\Theta _{\mu \nu}\, ,
\end{equation}
where 
\begin{equation}\nonumber
f_{R}\left( R,T\right) = {\partial f\left(R,T\right) \over \partial R} , \qquad
f_{T}\left( R,T\right) = {\partial f\left(R,T\right) \over \partial T} , \qquad
\frac{\delta \left(g^{\alpha \beta }T_{\alpha \beta }\right)}{\delta g^{\mu \nu}} =T_{\mu\nu}+\Theta _{\mu \nu} ,
\end{equation}
and $T_{\mu\nu}$ is the energy-momentum tensor which is defined as
\begin{equation}\label{em}
T_{\mu \nu }=-\frac{2}{\sqrt{-g}} \frac{\delta \left( \sqrt{-g}L_\mathrm{m}\right) }{\delta g^{\mu \nu}} 
                  = g_{\mu \nu} L_m - 2 {\delta L_m \over \delta g^{\mu \nu }}.
\end{equation}
The tensor $\Theta_{\mu \nu}$ is defined through the variation of the energy-momentum tensor with respect to the metric as
\begin{equation}\label{theta_tensor}
\Theta_{\mu \nu} \equiv  g^{\alpha \beta }\frac{\delta T_{\alpha \beta}}{\delta g^{\mu \nu}} 
                      = -2 T_{\mu\nu}  + g_{\mu\nu} L_m 
                                - 2 g^{\alpha\beta} \frac{\delta^2 L_m}{\delta g^{\mu \nu} \delta g^{\alpha \beta}}
\end{equation}

It is assumed that the function $f(R,T) = R + \eta T$, with constant $\eta = \lambda \kappa^2$. This choice for $f(R,T)$ is utilized for studying different cosmological topics \cite{Harko:2011kv,Moraes:2015uxq,Carvalho:2017pgk,Moraes:2017rrv,Moraes:2017mir,Moraes:2016akv,
Azizi:2012yv,Moraes:2014cxa,Moraes:2015kka,Reddy:2013bva}. Substituting it in the field equation \eqref{fieldequation} leads to
\begin{equation}\label{myfieldequation}
G_{\mu \nu} \equiv R_{\mu \nu} - {1 \over 2} g_{\mu \nu} R = \kappa^2 T_{\mu \nu} 
                             + \eta \left( {1 \over 2} g_{\mu \nu} T - T_{\mu \nu} - \Theta_{\mu \nu} \right).
\end{equation}

%%%%%%%%%%%%%%%%%%%%%%%%%%%%%%%%%%%%%%%%%%
%%%%%%%%%%%%%%%%%%%%%%%%%%%%%%%%%%%%%%%%%%
%%%%%%%%%%%%%%%%%%%%%%%%%%%%%%%%%%%%%%%%%%
%%%%%%%%%%%%%%%%%%%%%%%%%%%%%%%%%%%%%%%%%%
%%%%%%%%%%%%%%%%%%%%%%%%%%%%%%%%%%%%%%%%%%
%%%%%%%%%%%%%%%%%%%%%%%%%%%%%%%%%%%%%%%%%%
\subsection{Tachyon field} 
The matter field in the action \eqref{action} is taken as a tachyon field with the following Lagrangian
\begin{equation}\label{tachyon_lagrangian}
L_{t} = - V(\phi) \sqrt{1 + g^{\alpha\beta} \nabla_\alpha \phi \nabla_\beta \phi} \; .
\end{equation}
Then, the corresponding energy-momentum tensor is obtained as
\begin{equation}\label{tachyon_em}
T_{\mu\nu} = {V(\phi) \over \sqrt{1 + \nabla^\alpha \phi \nabla_\alpha \phi}} \; \nabla_\mu \phi \nabla_\nu \phi
                  - g_{\mu\nu} V(\phi) \sqrt{1 + \nabla^\alpha \phi \nabla_\alpha \phi}
\end{equation}
and for the tensor $\Theta_{\mu\nu}$, one has
\begin{eqnarray}\label{theta_tensor_tachyon}
\Theta_{\mu \nu} & = & -2 \; \left( {V(\phi) \over \sqrt{1 + \nabla^\alpha \phi \nabla_\alpha \phi}} \; 
                                      \nabla_\mu \phi \nabla_\nu \phi  
                         - g_{\mu\nu} V(\phi) \sqrt{1 + \nabla^\alpha \phi \nabla_\alpha \phi} \right)  \\  
   & & \qquad  + g_{\mu\nu} \; \left( - V(\phi) \sqrt{1 + g^{\alpha\beta} \nabla_\alpha \phi \nabla_\beta \phi} \right)
 - {1 \over 2} {V(\phi) \nabla_\alpha \phi \nabla^\alpha \phi \over \left(1 + \nabla^\lambda \phi \nabla_\lambda \phi \right)^{3/2}} \; \nabla_\mu \phi \nabla_\nu \phi . \nonumber 
\end{eqnarray}
Assuming that the geometry of the universe is described by a spatially flat FLRW metric, 
\begin{equation}
ds^2 = -dt^2 + a^2(t) \left( dx^2 + dy^2 + dz^2 \right).
\end{equation}
the Friedmann equations are obtained as
\begin{equation}\label{tachyon_FriedmannEq1_B}
H^2 = {\kappa^2 \over 3} \; {V(\phi) \over \sqrt{1 - \dot{\phi}^2} } \; 
              \left( \big( 1 + 2 \lambda \big) - {\lambda \over 2} \dot{\phi}^2 
                                      - {\lambda \over 2} \; {\dot{\phi}^4 \over (1 - \dot{\phi}^2)}\right)
\end{equation}
and
\begin{equation}\label{tachyon_FriedmannEq2}
-2\dot{H} - 3H^2 = \kappa^2 {V(\phi) \over \sqrt{1 - \dot{\phi}^2}}  \; \left( {-\lambda \over 2} \; 
                                    - \big( 1 + {3 \over 2}  \lambda \big) \;  \big(1 - \dot{\phi}^2 \big) \right).
\end{equation}
Combining these two equations gives the time derivative of the Hubble parameter as
\begin{equation}\label{tachyon_Hdot_B}
-2 \dot{H} = \kappa^2 \; {V(\phi) \dot{\phi}^2 \over \sqrt{1 - \dot{\phi}^2} } \; 
                           \left( \big( 1 + \lambda \big)   
                                      - {\lambda \over 2} \; {\dot{\phi}^2 \over (1 - \dot{\phi}^2)}\right) . 
\end{equation}
The equation of motion of the tachyon field is read from the last three equation, which is given by
\begin{eqnarray}
\left[  {(1 +  \lambda) \over (1 - \dot{\phi}^2)} 
                        - {3\lambda \over 2} \; {\dot{\phi}^2 \over (1 - \dot{\phi}^2)} 
                           - {3\lambda \over 2} \; {\dot{\phi}^4 \over (1 - \dot{\phi}^2)^2}\right] \ddot{\phi} 
                & + & 3H \dot{\phi} \left( \big( 1 + \lambda \big)   
                                      - {\lambda \over 2} \; {\dot{\phi}^2 \over (1 - \dot{\phi}^2)}\right) \\
            & & \qquad \qquad \qquad + \left( \big( 1 + 2 \lambda \big) - {\lambda \over 2} \dot{\phi}^2 
                   - {\lambda \over 2} \; {\dot{\phi}^4 \over (1 - \dot{\phi}^2)} \right) \; {V' \over V} = 0 . \nonumber    
\end{eqnarray}

We have derived the main dynamical equations of our model where the tachyon field plays the role of matter source. With these equations, we next are going to investigate the slow-roll inflation.

%%%%%%%%%%%%%%%%%%%%%%%%%%%%%%%%%%%%%%%%%%
%%%%%%%%%%%%%%%%%%%%%%%%%%%%%%%%%%%%%%%%%%
%%%%%%%%%%%%%%%%%%%%%%%%%%%%%%%%%%%%%%%%%%
%%%%%%%%%%%%%%%%%%%%%%%%%%%%%%%%%%%%%%%%%%
%%%%%%%%%%%%%%%%%%%%%%%%%%%%%%%%%%%%%%%%%%
%%%%%%%%%%%%%%%%%%%%%%%%%%%%%%%%%%%%%%%%%%
%%%%%%%%%%%%%%%%%%%%%%%%%%%%%%%%%%%%%%%%%%
%%%%%%%%%%%%%%%%%%%%%%%%%%%%%%%%%%%%%%%%%%
%%%%%%%%%%%%%%%%%%%%%%%%%%%%%%%%%%%%%%%%%%
\section{Slow-roll tachyon inflation}\label{tachyon_inflation}
One of the main characteristics of slow-roll inflation is the slow-roll approximations that are defined by a set of parameters, known as the slow-roll parameters. These parameters plays a crucial role in the scenario. It is assumed that, these parameters are smaller than one throughout the inflationary phase resulting to a major simplification to the dynamical equations of inflation. The first slow-roll parameter is
\begin{equation}\label{epsilon}
\epsilon_1 = {-\dot{H} \over H^2} = {3 \over 2}
        {
        \big( 1 + \lambda \big) \dot{\phi}^2   
                                      - {\lambda \over 2} \; {\dot{\phi}^4 \over (1 - \dot{\phi}^2)}
\over         
         \big( 1 + 2 \lambda \big) - {\lambda \over 2} \dot{\phi}^2 
                                      - {\lambda \over 2} \; {\dot{\phi}^4 \over (1 - \dot{\phi}^2)}
        } \ll 1 ,
\end{equation}
stating that the rate of the Hubble parameter during a Hubble time is smaller than one. Following the same statement about the time derivative of the tachyon field leads to definition of the second slow-roll parameter as
\begin{equation}\label{delta}
\delta = {\ddot{\phi} \over H \dot{\phi}}  \ll 1 .
\end{equation}
\textbf{The smallness of $\epsilon_1$, during inflation, indicates that the kinetic term, $\dot{\phi}^2$, is small as well. Consequently, throughout the inflationary period, it holds that $\dot{\phi}^4 \ll \dot{\phi}^2 \ll 1$.} It leads to this conclusion that $(1 - \dot{\phi}^2) \simeq 1$ leading to the simplification of the Friedmann equations.

%%%%%%%%%%%%%%%%%%%%%%%%%%%%%%%%%%%%%%%%%%
%%%%%%%%%%%%%%%%%%%%%%%%%%%%%%%%%%%%%%%%%%
%%%%%%%%%%%%%%%%%%%%%%%%%%%%%%%%%%%%%%%%%%
%%%%%%%%%%%%%%%%%%%%%%%%%%%%%%%%%%%%%%%%%%
%%%%%%%%%%%%%%%%%%%%%%%%%%%%%%%%%%%%%%%%%%
%%%%%%%%%%%%%%%%%%%%%%%%%%%%%%%%%%%%%%%%%%
\subsection{Equations under slow-roll approximations}
After applying the slow-roll approximations, the dynamical equations of the model gain a more simple form and are rewritten as
\begin{eqnarray}
H^2 & = & {\kappa^2 \over 3} \; \big( 1 + 2\lambda \big) \; V(\phi), \label{sr-friedmann} \\
\dot{H} & = & - {\kappa^2 \over 2} \; \big( 1 + \lambda \big) \; \dot{\phi}^2 \; V(\phi), \label{sr-dHt} \\
3 H \dot{\phi} & = & - \; {1 + 2\lambda \over 1 + \lambda} \; {V' \over V}. \label{sr-eom}
\end{eqnarray}
From Eqs.\eqref{sr-friedmann} and \eqref{sr-eom}, the time derivative of the tachyon field is obtained as
\begin{equation}\label{sr-dphi}
\dot{\phi}^2 = {1 \over 3 \kappa^2} \; {2\lambda + 1 \over (\lambda + 1)^2} \; {V'^2 \over V^3}.
\end{equation}
The above equations are utilized to express the slow-roll parameter $\epsilon_1$ in terms of the potential
\begin{equation}\label{sr-epsilon1}
\epsilon_1 = {1 \over 2 \kappa^2} \; {1 \over (1 + \lambda)} \; {V'^2 \over V^3}
                                 = {3 \over 2} \; {(1 + \lambda) \over (1 + 2\lambda)} \; \dot{\phi}^2.
\end{equation}
Besides the second slow-roll parameter $\delta$, which introduced in Eq.\eqref{delta} and is commonly used, there is a well-known hierarchy procedure for defining the slow-roll parameters as
\begin{equation}
\epsilon_{n+1} = {\dot{\epsilon}_n \over H \epsilon_n}, \qquad n \geq 1.
\end{equation}
Defining the slow-roll parameter through this procedure would be more efficient for us when we are going to work with the perturbation parameters. The second slow-roll parameter $\epsilon_2$ is derived as\footnote{Note that for this model, the slow-roll parameter $\epsilon_2$ is very close to the slow-roll parameter $\delta$; namely $\epsilon_2 = 2\ddot{\phi} / H \dot{\phi} = 2 \delta$. }
\begin{equation}\label{sr-epsilon2}
\epsilon_2 = {\dot{\epsilon}_1 \over H \epsilon_1} = {1 \over \kappa^2 (1 + \lambda)} \;
                                                                  \left( {3 V'^2 \over V^3} - 2{V'' \over V^2} \right).
\end{equation}

%%%%%%%%%%%%%%%%%%%%%%%%%%%%%%%%%%%%%%%%%%
%%%%%%%%%%%%%%%%%%%%%%%%%%%%%%%%%%%%%%%%%%
%%%%%%%%%%%%%%%%%%%%%%%%%%%%%%%%%%%%%%%%%%
%%%%%%%%%%%%%%%%%%%%%%%%%%%%%%%%%%%%%%%%%%
%%%%%%%%%%%%%%%%%%%%%%%%%%%%%%%%%%%%%%%%%%
%%%%%%%%%%%%%%%%%%%%%%%%%%%%%%%%%%%%%%%%%%
\subsection{Perturbations}
Quantum fluctuations are the minute variations in the universe's density that result from the quantum mechanical uncertainty principle. These fluctuations are believed to be the origin of all cosmological structure, including galaxies, clusters, and superclusters, according to the inflationary model of the universe. It is believed that during inflation, the cosmos expanded quickly, extending quantum fluctuations in matter and energy density to macroscopic sizes.
The earliest conditions for the emergence of structure in the cosmos were created by these fluctuations, which were then frozen into spacetime. Many observational evidences, like as the cosmic microwave background radiation, which exhibits temperature variations that are consistent with inflationary model predictions, provide support to this scenario.
Overall, quantum fluctuations are an important part of our understanding of the early universe and the evolution of the cosmos because they play a key role in the building of structure in the universe and have been validated by numerous studies.  \\

It is crucial to verify any new inflationary model using the available facts. In this regard, we introduce the key perturbation parameters for our model here, and we'll use them in the following section when we compare the model to data for various potential kinds. In terms of scalar perturbations, the amplitude of the perturbations is written as  \cite{Garriga:1999vw} \footnote{From \cite{Garriga:1999vw} and the shape of the action \eqref{action} and the Lagrangian \eqref{tachyon_lagrangian}, it is realized that the effective matter Lagrangian could in general be written as a function of $X = \dot{\phi}^2 / 2$ and the potential $V(\phi)$, in another word, the Lagrangian density of the matter field is $\mathcal{L}(\phi,X)$. Then, the model is a k-essence model and the quantum perturbations of the k-essence model is studied in \cite{Garriga:1999vw}. }
\begin{equation}\label{Ps}
\mathcal{P}_s  = {1 \over 4\pi^2} \; {H^4 \over c_s (\rho_{eff} + p_{eff})}
\end{equation}
where $c_s$ is the sound speed, defined as\footnote{In general, the sound speed is defined as $c_s^2 = {d p / d \rho}$. But, in our case, the energy density and pressure in the relation are not the energy density and pressure of the original tachyon field defined as $\rho = V(\phi) / \sqrt{1 - \dot{\phi}^2}$ and $p = V(\phi) \sqrt{1 - \dot{\phi}^2}$. In fact, they are the effective energy density and pressure which is defined from the right-hand side of Eqs.\eqref{eff_rho} and \eqref{eff_p}. For more detail refer to \cite{Garriga:1999vw}.} \cite{Garriga:1999vw}
\begin{equation}\label{sound_speed}
c_s^2 = { (1 + \lambda) - (1 + {3 \over 2} \lambda ) \dot{\phi}^2 
             \over 
                   (1+\lambda) + {\lambda \over 2} \dot{\phi}^2 
                            + {\lambda \over 2} { \dot{\phi}^4 - 4 \dot{\phi}^2 \over (1 - \dot{\phi}^2) }}.
\end{equation}
and
\begin{eqnarray}
  \rho_{eff} &=& {V(\phi) \over \sqrt{1 - \dot{\phi}^2} } \; 
                           \left( \big( 1 + 2 \lambda \big) - {\lambda \over 2} \dot{\phi}^2 
                                      - {\lambda \over 2} \; {\dot{\phi}^4 \over (1 - \dot{\phi}^2)}\right), \label{eff_rho}\\
  p_{eff} &=& {V(\phi) \over \sqrt{1 - \dot{\phi}^2}}  \; \left( {-\lambda \over 2} \; 
                                    - \big( 1 + {3 \over 2}  \lambda \big) \;  \big(1 - \dot{\phi}^2 \big) \right). \label{eff_p}
\end{eqnarray}
%with $\rho$ and $p$ are the energy density and pressure of the original tachyon field given by Eq.\eqref{energy_pressure}. \\
%Also, $X = \dot{\phi}^2/2$ and the constant parameter $\theta$ is defined as $\theta = (1 + 2\lambda) / 3(1+\lambda)$. \\
The scalar spectral index, which is defined through the amplitude of the scalar field, is given by
\begin{equation}
n_s = 1 - (2 \epsilon_1 + \epsilon_2 + s),
\end{equation}\label{ns}
where $s = \dot{c}_s / H c_s$ is another slow-roll parameter which determines the rate of the sound speed in a Hubble time.\\
Regarding the tensor perturbations, there is the tensor-to-scalar ratio, which is very essential in examining an inflationary model. The parameter is expressed as follows
\begin{equation}\label{r}
r = 16 \; c_s \; \epsilon_1 .
\end{equation}
In the following section, we are going to examine the model for some specific types of the potential and compare the results with data.

%%%%%%%%%%%%%%%%%%%%%%%%%%%%%%%%%%%%%%%%%%
%%%%%%%%%%%%%%%%%%%%%%%%%%%%%%%%%%%%%%%%%%
%%%%%%%%%%%%%%%%%%%%%%%%%%%%%%%%%%%%%%%%%%
%%%%%%%%%%%%%%%%%%%%%%%%%%%%%%%%%%%%%%%%%%
%%%%%%%%%%%%%%%%%%%%%%%%%%%%%%%%%%%%%%%%%%
%%%%%%%%%%%%%%%%%%%%%%%%%%%%%%%%%%%%%%%%%%
%%%%%%%%%%%%%%%%%%%%%%%%%%%%%%%%%%%%%%%%%%
%%%%%%%%%%%%%%%%%%%%%%%%%%%%%%%%%%%%%%%%%%
%%%%%%%%%%%%%%%%%%%%%%%%%%%%%%%%%%%%%%%%%%
\section{Typical examples}\label{examples}
In this section, three types of potentials as power-law, generalized T-mode, and the inverse coshyperbolic are taken, and the model is studied in detail for each case.

%%%%%%%%%%%%%%%%%%%%%%%%%%%%%%%%%%%%%%%%%%
%%%%%%%%%%%%%%%%%%%%%%%%%%%%%%%%%%%%%%%%%%
%%%%%%%%%%%%%%%%%%%%%%%%%%%%%%%%%%%%%%%%%%
%%%%%%%%%%%%%%%%%%%%%%%%%%%%%%%%%%%%%%%%%%
%%%%%%%%%%%%%%%%%%%%%%%%%%%%%%%%%%%%%%%%%%
%%%%%%%%%%%%%%%%%%%%%%%%%%%%%%%%%%%%%%%%%%
\subsection{power-law}
We begin with the most common potential, i.e. power-law potential
\begin{equation}\label{pot_powerlaw}
  V(\phi) = V_0 \; \phi^n .
\end{equation}
Utilizing the potential, the first and second slow-roll parameters are obtained as
\begin{eqnarray}\label{srp_powerlaw}
  \epsilon_1 &=& {n^2 \over 2\kappa^2 (1+\lambda) V_0} \; {1 \over \phi^{n+2}}, \\
  \epsilon_2 &=& {n(n+2) \over \kappa^2 (1+\lambda) V_0} \; {1 \over \phi^{n+2}}.
\end{eqnarray}
and the time derivative of the scalar field is derived by inserting the potential \eqref{pot_powerlaw} in \eqref{sr-dHt} as
\begin{equation}\label{dphit_powerlaw}
  \dot{\phi}^2 = {n^2 (1 + 2\lambda) \over 3\kappa^2 (1+\lambda)^2 V_0} \; {1 \over \phi^{n+2}}
\end{equation}
Inflation ends as the acceleration reaches zero, which is expressed by the relation $\epsilon_1 = 1$. Solving the relation, one can estimate the scalar field value at the end of inflation as
\begin{equation}\label{phi_end_powerlaw}
  \phi_e^{n+2} = {n^2 \over 2\kappa^2 (1+\lambda) V_0}.
\end{equation}
Following the definition of the number of e-folds
\begin{eqnarray}\label{efold_powerlaw}
  N & = & \int_{t_\star}^{t_e} H dt = \int_{\phi_\star}^{\phi_e} {H \over \dot{\phi}} \; d\phi \\
    & = & -\kappa^2 (1+\lambda) \int_{\phi_\star}^{\phi_e} {V^2 \over V'} \; d\phi
\end{eqnarray}
and solving the integral, the scalar field is written as a function of the number of e-fold as
\begin{equation}\label{phi_star_powerlaw}
  \phi_\star^{n+2} = {n^2 \over 2\kappa^2 (1+\lambda) V_0} \;
               \left( 1 + {2 (n+2) \over n} \; N \right).
\end{equation}
Using the potential and the obtained $\phi_\star$, and substituting them in the slow-roll parameters \eqref{sr-epsilon1} and \eqref{sr-epsilon2}, the scalar spectral index and the tensor-to-scalar ratio are achieved
\begin{eqnarray}\label{rns_powerlaw}
  n_s &=& 1 - s^\star - {4(n+1) \over n} \; \left( 1 + {2(n+2) \over n} \; N \right)^{-1} \\
  r &=& 16 \; c_s^\star \; \left( 1 + {2(n+2) \over 3n} \; N \right)^{-1} ,
\end{eqnarray}
where the sound speed is obtained by combining Eqs.\eqref{sound_speed} and \eqref{dphit_powerlaw}, and \eqref{phi_star_powerlaw}. \\
%\begin{equation*}
%  c_s = \sqrt{1 - {2 \over 3} {1+2\lambda \over 1+\lambda} \; \left( 1 + {2(n+1) \over 3n} \; N \right)^{-1}}
%\end{equation*}

%%%%%%%%%%%%%%%%%%%%%%%%%%%%%%%%%%%%%%%%%
\begin{figure}[t]
\centering
\subfigure[]{\includegraphics[width = 7cm]{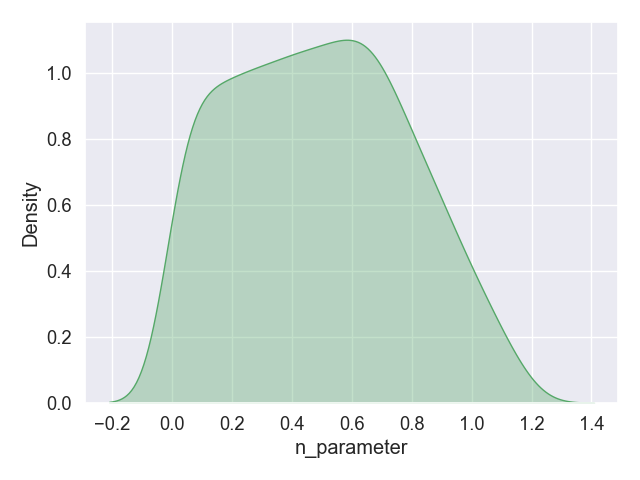}}
\subfigure[]{\includegraphics[width = 7cm]{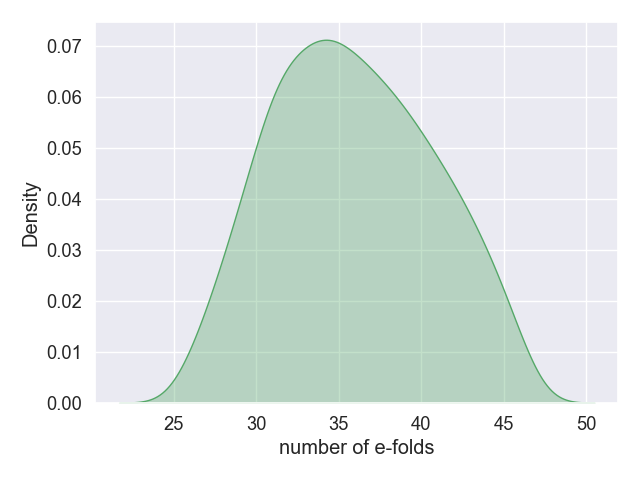}}
\caption{The density plot of the parameter $n$ and the number of e-fold $N$. }\label{powerlaw_densityplot}
\end{figure}
%%%%%%%%%%%%%%%%%%%%%%%%%%%%%%%%%%%%%%%%%%

The perturbation parameters $n_s$ and $r$ depend on the model parameters $n$ and the number of e-folds $N$. If the presented model is a valid candidate of the inflation, it is required to be consistent with the data. Then, the free parameters of the model should be adjusted to ensure this consistency. Using Eqs.\eqref{rns_powerlaw} on a Python code and applying the data, desire sets of values for the parameters are obtained. Fig.\ref{powerlaw_densityplot} illustrates the density plots of the preferred values of $n$ and $N$. The horizontal line presents the accepted values of the corresponding parameters and the vertical axis states the probability density of values. Therefore, aside the point that it shows the acceptable values, we also could find out about the values with the higher probabilities. Then, the region of plot with higher peak presents the region that includes the maximum data point. From the figure, it is realized that maximum of data are for $n = 0.25-0.70 $ and $N = 32-37$\footnote{At the first of the coding, we call all the required packages which are "numpy", "pandas", and "matplotlib.pyplot". Then, we defined all the required functions, such as scalar spectral index, amplitude of the scalar function, the tensor-to-scalar ratio. Then, by defining a proper range for the free parameters of the model, and using "loop" command and "if and else" command, we find all $(n, N)$ points (as in first case of the study), where for each of these point the outcomes of the model for $n_s$ and $r$ stand perfectly in the observational range. In the previous step, all the obtained values for $n$ and $N$ are placed in a separate numpy.array, say A and B. Then, using the density plot command, from the pandas package, we portray a density plot sets A and B. Via this plot, The higher picks of these plots determine the range where the maximum of the obtained value (for each parameter) are standing there. Next, using the scatter plot command, from the pandas package, we provide two scatter plot for the two free parameter where the difference is only in the coloring of the points; the first scattering plot is colored based on the values of $n_s$ and the second plot is colored based on the value of $r$.}.  \\

\begin{figure}[t]
\centering
\subfigure[]{\includegraphics[width = 7cm]{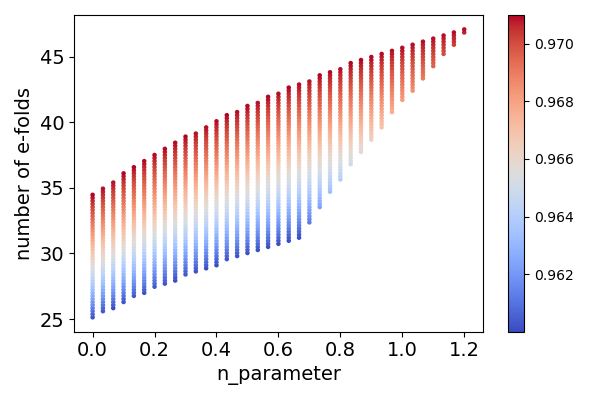}}
\subfigure[]{\includegraphics[width = 7cm]{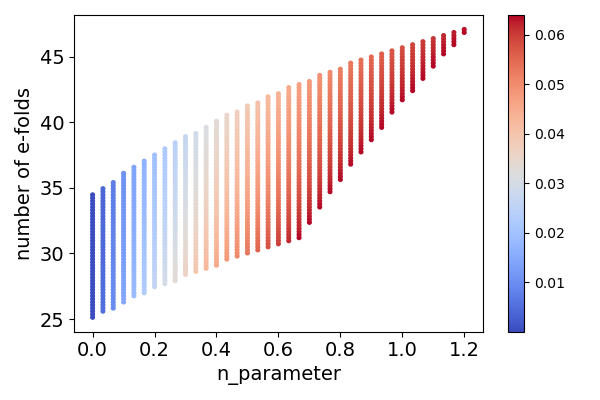}}
\caption{The figure displays the scatter plot of the parameter $n$ and the number of e-fold $N$ painted based on the values of a) scalar spectral index, and b) the tensor-to-scalar ratio.  }\label{powerlaw_rns}
\end{figure}

After obtaining the desired values of $n$ and $N$ from the density plots in Fig. 2, we need to consider the corresponding values of $n_s$ and $r$ for these parameters. This is displayed in Fig.\ref{powerlaw_rns} where scatter plots of the obtained values of $n$ and $N$ are presented, painted based on the values of $n_s$ and $r$. For the preferred values of the parameter $n$ and $N$, the scalar spectral index is about $n_s \approx 0.967$ and the tensor-to-scalar ratio stands around $r \approx 0.04$. \\
%%%%%%%%%%%%%%%%%%%%%%%%%%%%%%%%%%%%%%%%%%
%%%%%%%%%%%%%%%%%%%%%%%%%%%%%%%%%%%%%%%%%%

%%%%%%%%%%%%%%%%%%%%%%%%%%%%%%%%%%%
\begin{figure}[h]
\centering
\includegraphics[width = 7cm]{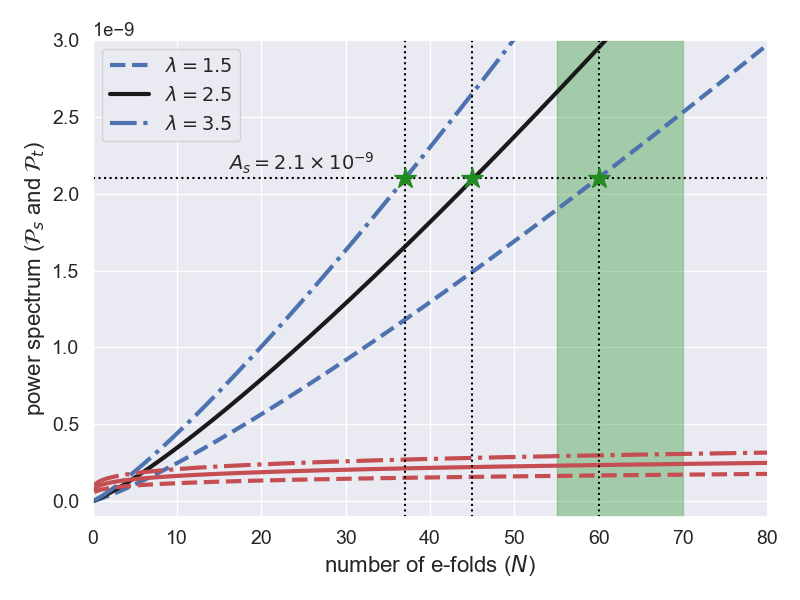}
\caption{The scalar and tensor power spectrum are plotted versus the number of e-folds. The scalar power spectrum are presented in blue color and the tensor power spectrum is in red. The power spectrum are plotted for $n = 0.5$ and three value of the constant $\lambda$ as: $\lambda = 2.5$ (solid line), $\lambda = 3.5$ (dashed line), and $\lambda = 1.5$ (dot-dashed line).}\label{powerlaw_powerspectrum}
\end{figure}
%%%%%%%%%%%%%%%%%%%%%%%%%%%%%%%%%%%%%

The power spectrum is an essential element in the studying inflationary models. The power spectrum of the scalar and tensor perturbations within the framework of the power-law potential in presented in Fig.\ref{powerlaw_powerspectrum}, providing an insight into the impact of the constant parameter $\lambda$. The power spectrum for the comoving modes within the Hubble radius at various stages of inflationary expansion is illustrated in the figure. The shaded green region is related to the comoving modes currently observed in the CMB, expecting to exit the horizon at the number of e-folds $N = 50 - 70$ (different values are mentioned in literature). The latest observational data implies that the amplitude of the scalar perturbations is approximately $2.1 \times 10^{-9}$, denoted by $A_s$ in the figure. Taking the constant $n = 0.5$, allowed by the Fig.\ref{powerlaw_rns}, it is determined that the curve crosses the $A_s$ line for $N = 47$. However, it is out of the our interest green region. To drag the collision point within the green region, one may pick a lower value for the constant parameter $\lambda$. For instance, with $\lambda = 1.5$ the curve cross the $A_s$ line at $N = 60$ e-folds. Nevertheless, this choice results in an unaccepted number of e-folds which leads to unfavorable values for the scalar spectral index and the tensor-to-scalar ratio.

%%%%%%%%%%%%%%%%%%%%%%%%%%%%%%%%%%%%%%%%%%
%%%%%%%%%%%%%%%%%%%%%%%%%%%%%%%%%%%%%%%%%%
%%%%%%%%%%%%%%%%%%%%%%%%%%%%%%%%%%%%%%%%%%
%%%%%%%%%%%%%%%%%%%%%%%%%%%%%%%%%%%%%%%%%%
%%%%%%%%%%%%%%%%%%%%%%%%%%%%%%%%%%%%%%%%%%
%%%%%%%%%%%%%%%%%%%%%%%%%%%%%%%%%%%%%%%%%%
\subsection{Generalized T-mode}
\begin{figure}[h]
\centering
\subfigure[]{\includegraphics[width = 7cm]{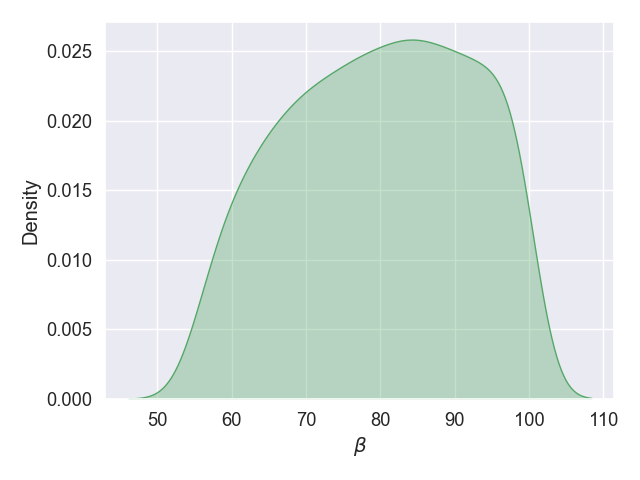}}
\subfigure[]{\includegraphics[width = 7cm]{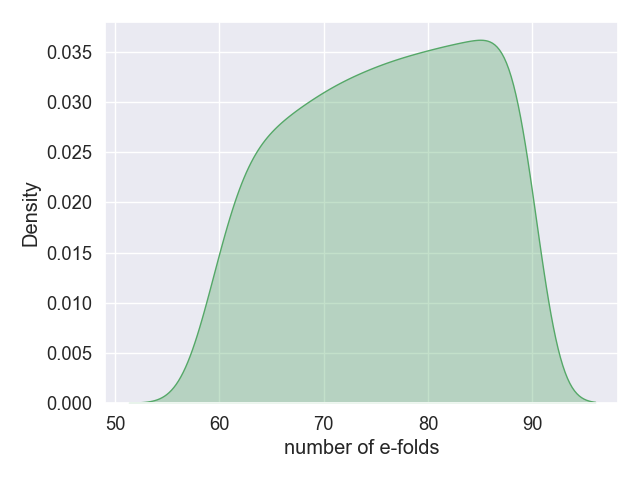}}
\caption{The density plot of the parameter $n$ and the number of e-fold $N$. }\label{GTmode_densityplot}
\end{figure}

For the second case, we are investigating a generalized T-mode potential
\begin{equation}\label{pot_gtmode}
  V(\phi) = V_0 \Big( 1 - \tanh^2(\alpha \phi ) \Big).
\end{equation}
Inserting the potential in Eqs.\eqref{sr-epsilon1} and \eqref{sr-epsilon2}, the slow-roll parameters are acquired as
\begin{eqnarray}\label{srp_gtmode}
  \epsilon_1 & = & {2 \alpha^2 \over \kappa^2 (1+\lambda) V_0} \;
                       { \tanh^2(\alpha \phi ) \over 1 - \tanh^2(\alpha \phi )}, \nonumber \\
  \epsilon_2 & = & {4 \alpha^2 \over \kappa^2 (1+\lambda) V_0} \;
                    {1 \over 1 - \tanh^2(\alpha \phi )}
\end{eqnarray}
and $\dot{\phi}$ is read from Eq.\eqref{sr-dphi}
\begin{equation}\label{dphi_gtmode}
  \dot{\phi}^2 = {4 \alpha^2 (1+2\lambda) \over 3 \kappa^2 (1+\lambda)^2 V_0} \;
                            { \tanh^2(\alpha \phi ) \over 1 - \tanh^2(\alpha \phi )}
\end{equation}
To estimate the scalar field at the end of inflation, it is required to solve the relation $\epsilon_1 = 1$. This leads to a second order equation, and the solution is given by
\begin{equation}\label{phi_end_gtmode}
  \tanh^2(\alpha \phi_e) = {\beta \over 1 + \beta}
\end{equation}
where the defined constant $\beta$ is
\begin{equation*}
  \beta \equiv {\kappa^2 (1+\lambda) V_0 \over 2 \alpha^2}.
\end{equation*}
By integrating the e-folds equation and after doing some manipulation, the scalar field at horizon crossing is derived
\begin{equation}\label{phi_star_gtmode}
  \tanh^2(\alpha \phi_\star) = {\beta \over 1+ \beta} \; \exp\left( {-2 N \over \beta} \right).
\end{equation}
Substituting the above result in the slow-roll parameters \eqref{srp_gtmode} and apply it into Eqs.\eqref{ns} and \eqref{r}, the scalar spectral index and the tensor-to-scalar ratio are obtained in terms of the free parameters $\beta$, $\lambda$ and the e-fold $N$.
\begin{eqnarray}
  ns &=& 1 - s^\star - {2 \over \beta} \; {(1+\beta) + \beta e^{-2N/\beta} \over \Big( (1+\beta) - \beta e^{-2N/\beta} \Big)} \label{gtmode_ns}\\
  r &=& 16 \; c_s^\star \;
           {e^{-2N/\beta} \over (1+\beta) - \beta e^{-2N/\beta}}, \label{gtmode_r}
\end{eqnarray}
where the sound speed at horizon crossing is read from Eqs.\eqref{sound_speed}, \eqref{dphi_gtmode}, and \eqref{phi_star_gtmode}. \\
%\begin{equation*}
%  cs^2 = 1 - {2(1 + 2\lambda) \over 3 (1 + \lambda)} \;
%             {e^{-2N/3\beta} \over (1+\beta) - \beta e^{-2N/3\beta}} \label{gtmode_r}
%\end{equation*}
The parameters $\beta$ and $N$ are the main parameters that determine the values of the scalar spectral index and the tensor-to-scalar ratio. Applying data on the programming, one could obtain the desired values for the parameters $n$ and $N$. Fig.\ref{GTmode_densityplot} illustrates the density plot for these two parameters for which they bring the model to a good consistency with the data. Since the vertical axis specifies the probability density of the values of the parameter. In this regard, for $\beta = 75-95$ and $N = 70-85$ are the values with higher probability density, however, there are notable data supporting the number of e-folds between $N = 60-70$ which is our range of interest range. \\ 
%%%%%%%%%%%%%%%%%%%%%%%%%%%%%%%%%%%%%%%%%%
%%%%%%%%%%%%%%%%%%%%%%%%%%%%%%%%%%%%%%%%%%

\begin{figure}[h]
\centering
\subfigure[]{\includegraphics[width = 7cm]{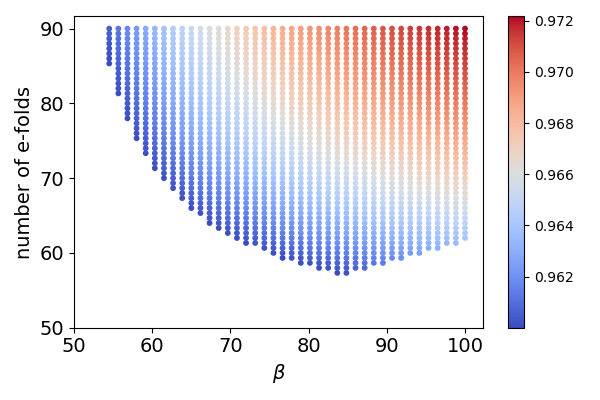}}
\subfigure[]{\includegraphics[width = 7cm]{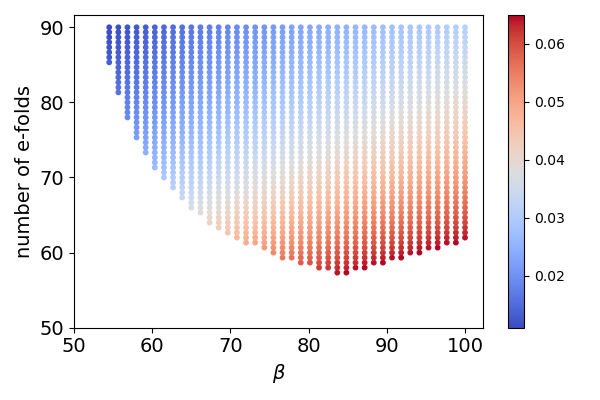}}
\caption{The figure displays the scatter plot of the parameter $n$ and the number of e-fold $N$ painted based on the values of a) scalar spectral index, and b) the tensor-to-scalar ratio.  }\label{GTmode_rns}
\end{figure}

The values of the scalar spectral index and the tensor-to-scalar for any pair and the parameters $\beta$ and $N$ are depicted in Fig.\ref{GTmode_rns}. Fron the figure, it is found out that for the mentioned desired values of the parameters, the scalar spectral index is about $n_s = 0.965 $ and the tensor-to-scalar is obtained as $r = 0.045$. These results perfectly stands in the observational range. \\
%%%%%%%%%%%%%%%%%%%%%%%%%%%%%%%%%%%%%%%%%%
%%%%%%%%%%%%%%%%%%%%%%%%%%%%%%%%%%%%%%%%%%

\begin{figure}[h]
\centering
\includegraphics[width = 7cm]{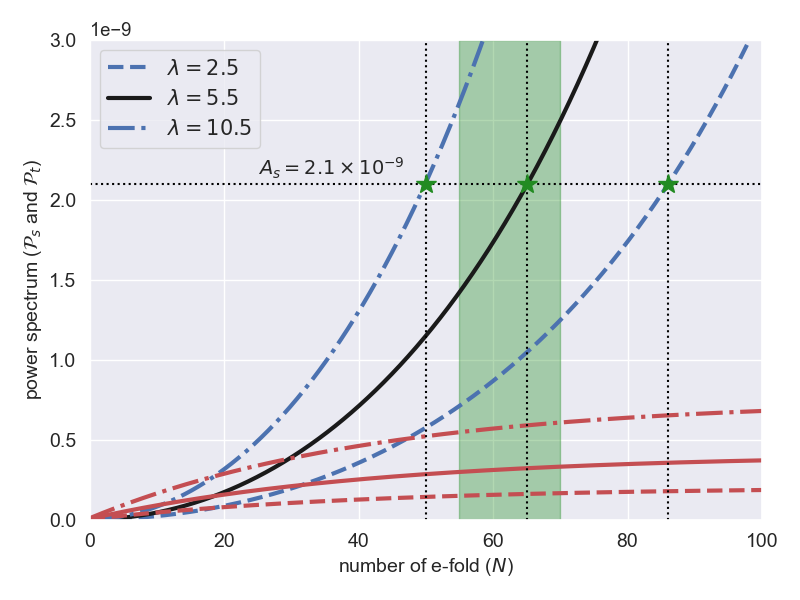}
\caption{The scalar and tensor power spectrum are plotted versus the number of e-folds. The scalar power spectrum are presented in blue color and the tensor power spectrum is in red. The power spectrum are plotted for $\beta = 85$ and three value of the constant $\lambda$ as: $\lambda = 5.5$ (solid line), $\lambda = 2.5$ (dashed line), and $\lambda = 10.5$ (dot-dashed line).}\label{GTmode_powerspectrum}
\end{figure}

The scalar and tensor power spectra for the comoving modes within the Hubble radius are depicted in Fig.\ref{GTmode_powerspectrum}. The observed CMB modes is believed are the modes which exited the Hubble radius at the number of e-folds about $N = 55-70$, as indicated by the green region in the figure. The modes corresponding to the higher/lower values of $N$ are associated with re-entry into the horizon in the future/past cosmological epoch. By selecting the constant value of $\beta = 85$, as allowed in Fig.\ref{GTmode_rns}, it is found that the $\mathcal{P}_s$ curve meet the $A_s$ line at $N = 65$, which is perfectly within our interest green region of the number of e-fold. It is noteworthy that this value of $N$ is confirmed by the constraints derived in Fig.\ref{GTmode_rns}, as the permissible range for the number of e-folds to get desirable values for the scalar spectral index and the tensor-to-scalar ratio. \\

%%%%%%%%%%%%%%%%%%%%%%%%%%%%%%%%%%%
%%%%%%%%%%%%%%%%%%%%%%%%%%%%%%%%%%%%%

A brief result of the model is presented in Table.\ref{GTmode_table}, where one could find the values for the scalar spectral index, the tensor-to-scalar ratio, the constant $V_0$, and the energy scale of inflation for different values of $\beta$ and $N$ selected from Fig.\ref{GTmode_densityplot}.
%%%5%%%%%%%%%%%%%%%%%%%%%%%%%%%%%%%%%%%%%%%
\begin{table}
\caption{The table present a brief results of the case for different values of $\beta$, $\lambda$, and the number of e-folds $N$. \label{GTmode_table}}
\centering
\begin{tabular}{p{1.3cm}p{1cm}p{1cm}p{1.5cm}p{1.5cm}p{2.5cm}p{2.5cm}p{2.5cm}}
\toprule
  $\lambda$ &    $\beta$ &    $N$ &   \quad   $n_s$ &  \quad    $r$ &    \qquad $V_0$ & \qquad  $\alpha$ & \qquad $ES$ \\
\midrule
    $1.0$ & $65$ & $60$ & $0.9579$ & $0.0453$ & $5.53 \times 10^{-10}$ & $2.92 \times 10^{-6}$ &  $4.65 \times 10^{-3}$ \\
    $1.0$ & $65$ & $65$ & $0.9597$ & $0.0378$ & $4.50 \times 10^{-10}$ & $2.63 \times 10^{-6}$ &  $4.44 \times 10^{-3}$ \\
    $1.0$ & $65$ & $70$ & $0.9613$ & $0.0317$ & $3.70 \times 10^{-10}$ & $2.38 \times 10^{-6}$ &  $4.25 \times 10^{-3}$ \\
    $1.0$ & $80$ & $60$ & $0.9608$ & $0.0565$ & $7.47 \times 10^{-10}$ & $3.06 \times 10^{-6}$ &  $4.91 \times 10^{-3}$ \\
    $1.0$ & $80$ & $65$ & $0.9629$ & $0.0483$ & $6.18 \times 10^{-10}$ & $2.78 \times 10^{-6}$ &  $4.72 \times 10^{-3}$ \\
    $1.0$ & $80$ & $70$ & $0.9646$ & $0.0414$ & $5.16 \times 10^{-10}$ & $2.54 \times 10^{-6}$ &  $4.55 \times 10^{-3}$ \\
    $1.0$ & $95$ & $60$ & $0.9626$ & $0.0654$ & $9.36 \times 10^{-10}$ & $3.14 \times 10^{-6}$ &  $5.10 \times 10^{-3}$ \\
    $1.0$ & $95$ & $65$ & $0.9647$ & $0.0566$ & $7.81 \times 10^{-10}$ & $2.87 \times 10^{-6}$ &  $4.92 \times 10^{-3}$ \\
    $1.0$ & $95$ & $70$ & $0.9666$ & $0.0493$ & $6.58 \times 10^{-10}$ & $2.63 \times 10^{-6}$ &  $4.75 \times 10^{-3}$ \\
  $100.0$ & $65$ & $60$ & $0.9579$ & $0.0453$ & $8.26 \times 10^{-12}$ & $2.53 \times 10^{-6}$ &  $1.63 \times 10^{-3}$ \\
  $100.0$ & $65$ & $65$ & $0.9598$ & $0.0379$ & $6.72 \times 10^{-12}$ & $2.29 \times 10^{-6}$ &  $1.55 \times 10^{-3}$ \\
  $100.0$ & $65$ & $70$ & $0.9613$ & $0.0318$ & $5.52 \times 10^{-12}$ & $2.07 \times 10^{-6}$ &  $1.49 \times 10^{-3}$ \\
  $100.0$ & $80$ & $60$ & $0.9609$ & $0.0565$ & $1.12 \times 10^{-11}$ & $2.65 \times 10^{-6}$ &  $1.72 \times 10^{-3}$ \\
  $100.0$ & $80$ & $65$ & $0.9629$ & $0.0483$ & $9.23 \times 10^{-12}$ & $2.41 \times 10^{-6}$ &  $1.65 \times 10^{-3}$ \\
  $100.0$ & $80$ & $70$ & $0.9646$ & $0.0414$ & $7.70 \times 10^{-12}$ & $2.20 \times 10^{-6}$ &  $1.59 \times 10^{-3}$ \\
  $100.0$ & $95$ & $60$ & $0.9626$ & $0.0654$ & $1.40 \times 10^{-11}$ & $2.73 \times 10^{-6}$ &  $1.78 \times 10^{-3}$ \\
  $100.0$ & $95$ & $65$ & $0.9648$ & $0.0567$ & $1.17 \times 10^{-11}$ & $2.49 \times 10^{-6}$ &  $1.72 \times 10^{-3}$ \\
  $100.0$ & $95$ & $70$ & $0.9666$ & $0.0494$ & $9.83 \times 10^{-12}$ & $2.29 \times 10^{-6}$ &  $1.66 \times 10^{-3}$ \\
$10000.0$ & $65$ & $60$ & $0.9579$ & $0.0453$ & $8.30 \times 10^{-14}$ & $2.53 \times 10^{-6}$ &  $5.15 \times 10^{-4}$ \\
$10000.0$ & $65$ & $65$ & $0.9598$ & $0.0379$ & $6.76 \times 10^{-14}$ & $2.28 \times 10^{-6}$ &  $4.92 \times 10^{-4}$ \\
$10000.0$ & $65$ & $70$ & $0.9613$ & $0.0318$ & $5.55 \times 10^{-14}$ & $2.07 \times 10^{-6}$ &  $4.71 \times 10^{-4}$ \\
$10000.0$ & $80$ & $60$ & $0.9609$ & $0.0565$ & $1.12 \times 10^{-13}$ & $2.65 \times 10^{-6}$ &  $5.44 \times 10^{-4}$ \\
$10000.0$ & $80$ & $65$ & $0.9629$ & $0.0483$ & $9.27 \times 10^{-14}$ & $2.41 \times 10^{-6}$ &  $5.23 \times 10^{-4}$ \\
$10000.0$ & $80$ & $70$ & $0.9646$ & $0.0414$ & $7.74 \times 10^{-14}$ & $2.20 \times 10^{-6}$ &  $5.03 \times 10^{-4}$ \\
$10000.0$ & $95$ & $60$ & $0.9626$ & $0.0654$ & $1.41 \times 10^{-13}$ & $2.72 \times 10^{-6}$ &  $5.64 \times 10^{-4}$ \\
$10000.0$ & $95$ & $65$ & $0.9648$ & $0.0567$ & $1.17 \times 10^{-13}$ & $2.48 \times 10^{-6}$ &  $5.44 \times 10^{-4}$ \\
$10000.0$ & $95$ & $70$ & $0.9666$ & $0.0494$ & $9.88 \times 10^{-14}$ & $2.28 \times 10^{-6}$ &  $5.26 \times 10^{-4}$ \\
\bottomrule
\end{tabular}
\end{table}
%%%%%%%%%%%%%%%%%%%%%%%%%%%%%%%%%%%%%%%%%%%%

%%%%%%%%%%%%%%%%%%%%%%%%%%%%%%%%%%%%%%%%%%
%%%%%%%%%%%%%%%%%%%%%%%%%%%%%%%%%%%%%%%%%%
%%%%%%%%%%%%%%%%%%%%%%%%%%%%%%%%%%%%%%%%%%
%%%%%%%%%%%%%%%%%%%%%%%%%%%%%%%%%%%%%%%%%%
%%%%%%%%%%%%%%%%%%%%%%%%%%%%%%%%%%%%%%%%%%
%%%%%%%%%%%%%%%%%%%%%%%%%%%%%%%%%%%%%%%%%%
\subsection{Inverse hyperbolic cosine potential}
As the last case, we consider the inverse hyperbolic cosine potential \cite{Leblond:2003db,Kim:2003he,Maloney:2003ck,Steer:2003yu}, given by
\begin{equation}\label{coshpot}
V(\phi) = {V_0 \over \cosh(\alpha \phi)}.
\end{equation}
where $\alpha$ and $V_0$ are two constants that will be determined later. Substituting the potential in Eqs.\eqref{sr-epsilon1} and \eqref{sr-epsilon2}, the first and second slow-roll parameters are obtained as
\begin{eqnarray}
\epsilon_1 & = & {\alpha^2 \over 2\kappa^2 (1 + \lambda) V_0} \; {\sinh^2(\alpha \phi) \over \cosh(\alpha \phi)}, \label{cosh_epsilon1} \\
\epsilon_2 & = & {\alpha^2 \over \kappa^2 (1 + \lambda) V_0} \; {\cosh^2(\alpha \phi) + 1 \over \cosh(\alpha \phi)},  \label{cosh_epsilon2}
\end{eqnarray}
and from Eq.\eqref{sr-dphi}, the time derivative of the tachyon field is
\begin{equation}
\dot{\phi}^2 = {\alpha^2 (1 + 2\lambda) \over 3 \kappa^2 (1 + \lambda)^2 V_0} \; {\sinh^2(\alpha \phi) \over \cosh(\alpha \phi)}.
\end{equation}
To estimate the parameters at the time of the horizon crossing, we first need to compute the field at the end of inflation. It is done by solving the relation $\epsilon_1 = 1$ that implies the time when the universe acceleration vanishes. It leads to
\begin{equation}\label{phi_end_cosh}
\sinh^2(\alpha \phi_e) = {\beta^2 \over 2} \left( 1 + \sqrt{1 + {4 \over \beta^2}} \right),
\end{equation}
where the constant $\beta$ is defined for more simplicity and is read as
\begin{equation}
\beta  \equiv {2\kappa^2 (1 + \lambda) V_0 \over \alpha^2}.
\end{equation}
Then, through the number of e-folds relation and using the above result, the field is read during inflation as
\begin{equation}\label{phi_star_cosh}
\tanh\left( {\alpha \phi_\star \over 2}\right) = \tanh\left( {\alpha \phi_e \over 2}\right) \; \exp\left( {-2 \over \beta} \; N \right).
\end{equation}
Applying the above tachyon field on the equations
\begin{eqnarray}
n_s & = & 1 - s^\star -  {4 \over \beta} \; \cosh(\alpha \phi_\star), \label{ns_cosh} \\
r & = & {16 \over \beta} \; c_s^\star \; {\sinh^2(\alpha \phi_\star) \over \cosh(\alpha \phi_\star)}. \label{r_cosh}
\end{eqnarray}

%%%%%%%%%%%%%%%%%%%%%%%%%%%%%%%%%%%%%%%%%%
\begin{figure}[ht]
\centering
\subfigure[]{\includegraphics[width = 7cm]{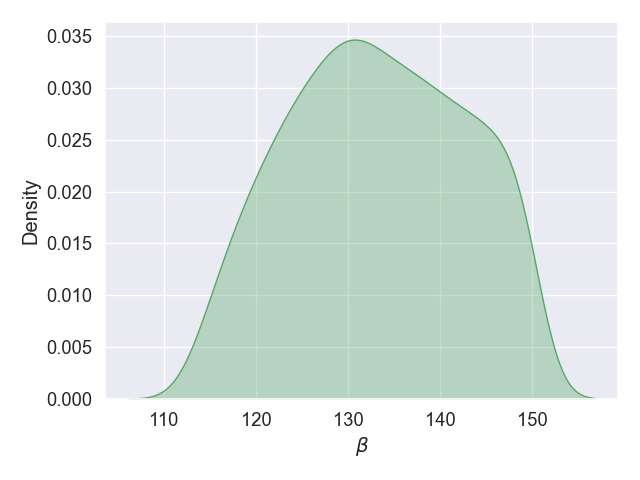}}
\subfigure[]{\includegraphics[width = 7cm]{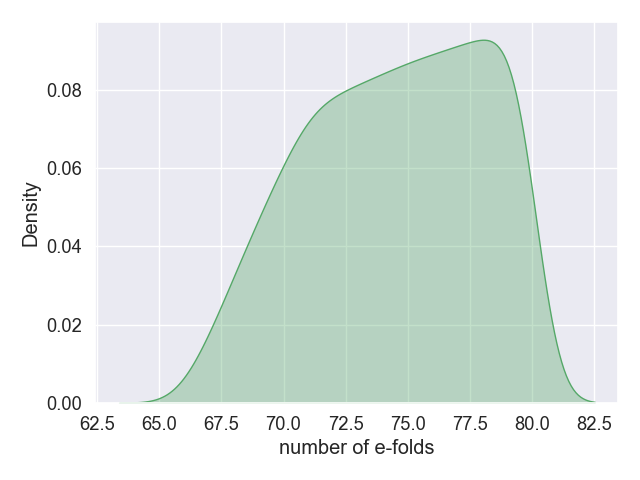}}
\caption{The density plot of the parameter $b$ and the number of e-fold $N$. }\label{Cosh_densityplot}
\end{figure}
%%%%%%%%%%%%%%%%%%%%%%%%%%%%%%%%%%%%%%%%%%

Applying Eq.\eqref{phi_star_cosh} on \eqref{ns_cosh} and \eqref{r_cosh}, it is realized that one needs to determine the parameters $\lambda$ and $N$. Utilizing the data and performing the same programming, two sets of values are obtained for the $\lambda$ and $N$ for which the model comes to a good agreement with the data. The density plot of the two parameters are displayed in Fig.\ref{Cosh_densityplot}, where one can find which values of the parameters are acceptable and also which values of the parameters have higher probabilities. From the figure one finds that the values $\beta = 125-140$ and $N = 72-78$ have higher probability. However, the plot shows that there are still some data supporting the number of e-folds at the range of $N = 65-70$.   \\

\begin{figure}[h]
\centering
\subfigure[]{\includegraphics[width = 7cm]{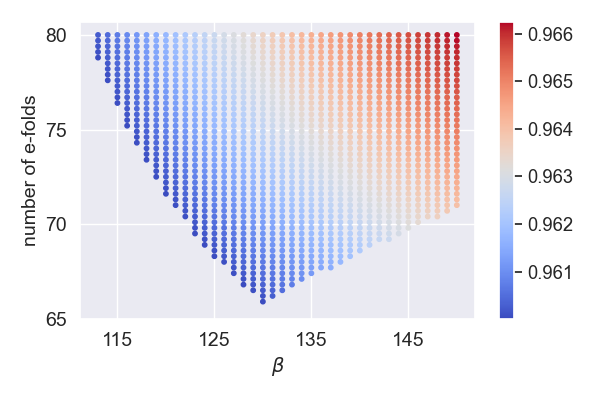}}
\subfigure[]{\includegraphics[width = 7cm]{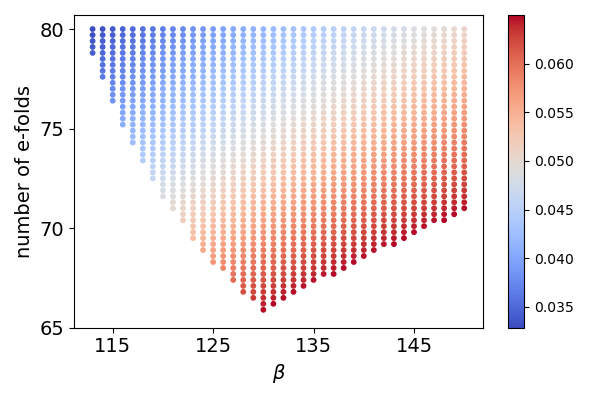}}
\caption{The figure displays the scatter plot of the parameter $b$ and the number of e-fold $N$ painted based on the values of a) scalar spectral index, and b) the tensor-to-scalar ratio.  }\label{Cosh_rns}
\end{figure}

Fig.\ref{Cosh_rns} presents the corresponding values of the scalar spectral index and the tensor-to-scalar ratio for the parameters above. For these preferred values of the parameters, the scalar spectral index and the tensor-to-scalar ratio are estimated as $n_s = 0.961$ and $r = 0.063$.
%%%%%%%%%%%%%%%%%%%%%%%%%%%%%%%%%%%%%%%%%%
%%%%%%%%%%%%%%%%%%%%%%%%%%%%%%%%%%%%%%%%%%

Asides from the last two parameters, we have two other free constants that are required to be determined. Using the Friedmann equation \eqref{sr-friedmann}, the defined potential, and by computing the equation at the time of the horizon crossing, one could estimate a valid value for the constant $V_0$ of the model as
\begin{equation}\label{V0_cosh}
V_0 = {24 \pi^2 \over \kappa^2 (1+2\lambda)} \; c_s \; \epsilon_1^\star \; \mathcal{P}_s^\star \; \cosh(\alpha \phi_\star).
\end{equation}
The constant $\alpha$ is also specified from the definition of $\beta$. 

%%%%%%%%%%%%%%%%%%%%%%%%%%%%%%%%%%%
\begin{figure}[h]
\centering
\includegraphics[width = 7cm]{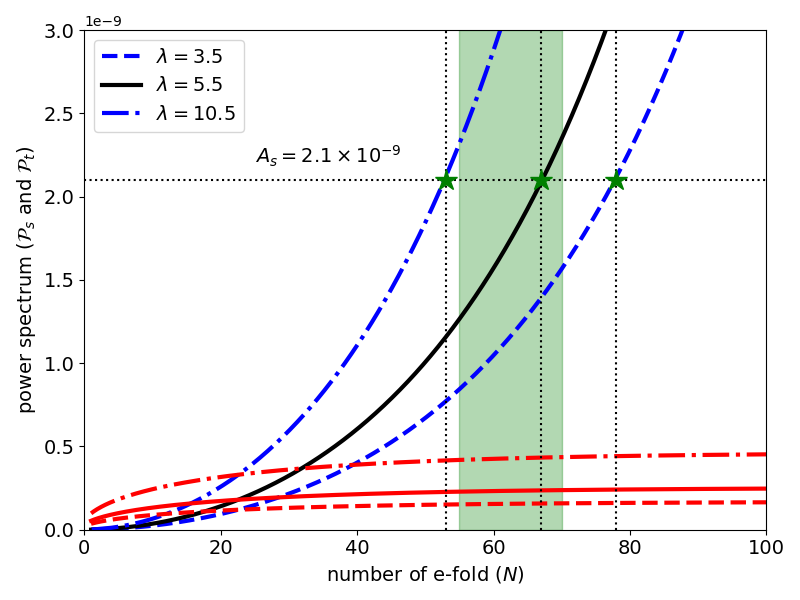}
\caption{The scalar and tensor power spectrum are plotted versus the number of e-folds. The scalar power spectrum are presented in blue color and the tensor power spectrum is in red. The power spectrum are plotted for $\beta = 130$ and three value of the constant $\lambda$ as: $\lambda = 5.5$ (solid line), $\lambda = 3.5$ (dashed line), and $\lambda = 10.5$ (dot-dashed line).}\label{Cosh_powerspectrum}
\end{figure}
%%%%%%%%%%%%%%%%%%%%%%%%%%%%%%%%%%%%%

We now turn our attention to determining whether the scalar power spectrum aligns with the amplitude $A_s$ appropriately in terms of the number of e-folds. The consideration is presented in Fig.\ref{Cosh_powerspectrum}, where both the scalar and the tensor power spectra are plotted as functions of the number of e-folds. One realizes that the scalar power spectrum meets the $A_s$ line within the shaded green region, at $N = 65$. It is verified that this value of $N$ falls within the permissible range as plotted in Figs.\ref{Cosh_densityplot} and \ref{GTmode_rns}. Therefore, the model could be taken as a proper model of inflation. In this case, we are going to determined the other free parameters of the model. \\

A brief result of the model is presented in Table.\ref{cosh_table}, where one could find the values for the scalar spectral index, the tensor-to-scalar ratio, the constant $V_0$, and the energy scale of inflation for different values of $\beta$ and $N$ selected from Fig.\ref{Cosh_densityplot}.
%%%%%%%%%%%%%%%%%%%%%%%%%%%%%%%%%
\begin{table}[h]
\caption{The table present a brief results of the case for different values of $\beta$ and $\lambda$ for the number of e-folds $N$. }\label{cosh_table}
\begin{tabular}{p{1.3cm}p{1cm}p{1cm}p{1.5cm}p{1.5cm}p{2.5cm}p{2.5cm}p{2.5cm}}
%\hline
\toprule
    $\lambda$ &  $\beta$ &  $N$ & \quad $n_s$ & \quad $r$ & \qquad $V_0$ & \qquad $\alpha$ & \qquad $ES$ \\
%\hline
\midrule
    $1.0$ & $125$ & $65$ & $0.9590$ & $0.0639$ & $8.43 \times 10^{-10}$ & $5.19 \times 10^{-6}$ & $5.07 \times 10^{-3}$ \\
    $1.0$ & $125$ & $67$ & $0.9596$ & $0.0598$ & $7.77 \times 10^{-10}$ & $4.99 \times 10^{-6}$ & $4.98 \times 10^{-3}$ \\
    $1.0$ & $125$ & $70$ & $0.9605$ & $0.0542$ & $6.90 \times 10^{-10}$ & $4.70 \times 10^{-6}$ & $4.86 \times 10^{-3}$ \\
    $1.0$ & $130$ & $65$ & $0.9597$ & $0.0667$ & $9.00 \times 10^{-10}$ & $5.26 \times 10^{-6}$ & $5.12 \times 10^{-3}$ \\
    $1.0$ & $130$ & $67$ & $0.9604$ & $0.0626$ & $8.31 \times 10^{-10}$ & $5.06 \times 10^{-6}$ & $5.04 \times 10^{-3}$ \\
    $1.0$ & $130$ & $70$ & $0.9613$ & $0.0569$ & $7.39 \times 10^{-10}$ & $4.77 \times 10^{-6}$ & $4.92 \times 10^{-3}$ \\
    $1.0$ & $135$ & $65$ & $0.9604$ & $0.0694$ & $9.56 \times 10^{-10}$ & $5.32 \times 10^{-6}$ & $5.17 \times 10^{-3}$ \\
    $1.0$ & $135$ & $67$ & $0.9611$ & $0.0653$ & $8.84 \times 10^{-10}$ & $5.12 \times 10^{-6}$ & $5.09 \times 10^{-3}$ \\
    $1.0$ & $135$ & $70$ & $0.9620$ & $0.0596$ & $7.88 \times 10^{-10}$ & $4.83 \times 10^{-6}$ & $4.98 \times 10^{-3}$ \\
  $100.0$ & $125$ & $65$ & $0.9590$ & $0.0639$ & $1.26 \times 10^{-11}$ & $4.51 \times 10^{-6}$ & $1.77 \times 10^{-3}$ \\
  $100.0$ & $125$ & $67$ & $0.9597$ & $0.0598$ & $1.16 \times 10^{-11}$ & $4.33 \times 10^{-6}$ & $1.74 \times 10^{-3}$ \\
  $100.0$ & $125$ & $70$ & $0.9605$ & $0.0542$ & $1.03 \times 10^{-11}$ & $4.08 \times 10^{-6}$ & $1.70 \times 10^{-3}$ \\
  $100.0$ & $130$ & $65$ & $0.9598$ & $0.0668$ & $1.34 \times 10^{-11}$ & $4.57 \times 10^{-6}$ & $1.79 \times 10^{-3}$ \\
  $100.0$ & $130$ & $67$ & $0.9604$ & $0.0627$ & $1.24 \times 10^{-11}$ & $4.39 \times 10^{-6}$ & $1.76 \times 10^{-3}$ \\
  $100.0$ & $130$ & $70$ & $0.9613$ & $0.0570$ & $1.10 \times 10^{-11}$ & $4.14 \times 10^{-6}$ & $1.72 \times 10^{-3}$ \\
  $100.0$ & $135$ & $65$ & $0.9604$ & $0.0695$ & $1.43 \times 10^{-11}$ & $4.62 \times 10^{-6}$ & $1.81 \times 10^{-3}$ \\
  $100.0$ & $135$ & $67$ & $0.9611$ & $0.0654$ & $1.32 \times 10^{-11}$ & $4.45 \times 10^{-6}$ & $1.78 \times 10^{-3}$ \\
  $100.0$ & $135$ & $70$ & $0.9620$ & $0.0596$ & $1.18 \times 10^{-11}$ & $4.20 \times 10^{-6}$ & $1.74 \times 10^{-3}$ \\
$10000.0$ & $125$ & $65$ & $0.9590$ & $0.0639$ & $1.27 \times 10^{-13}$ & $4.50 \times 10^{-6}$ & $5.61 \times 10^{-4}$ \\
$10000.0$ & $125$ & $67$ & $0.9597$ & $0.0598$ & $1.17 \times 10^{-13}$ & $4.32 \times 10^{-6}$ & $5.52 \times 10^{-4}$ \\
$10000.0$ & $125$ & $70$ & $0.9605$ & $0.0542$ & $1.04 \times 10^{-13}$ & $4.07 \times 10^{-6}$ & $5.38 \times 10^{-4}$ \\
$10000.0$ & $130$ & $65$ & $0.9598$ & $0.0668$ & $1.35 \times 10^{-13}$ & $4.56 \times 10^{-6}$ & $5.67 \times 10^{-4}$ \\
$10000.0$ & $130$ & $67$ & $0.9604$ & $0.0627$ & $1.25 \times 10^{-13}$ & $4.38 \times 10^{-6}$ & $5.58 \times 10^{-4}$ \\
$10000.0$ & $130$ & $70$ & $0.9613$ & $0.0570$ & $1.11 \times 10^{-13}$ & $4.13 \times 10^{-6}$ & $5.45 \times 10^{-4}$ \\
$10000.0$ & $135$ & $65$ & $0.9604$ & $0.0695$ & $1.44 \times 10^{-13}$ & $4.61 \times 10^{-6}$ & $5.73 \times 10^{-4}$ \\
$10000.0$ & $135$ & $67$ & $0.9611$ & $0.0654$ & $1.33 \times 10^{-13}$ & $4.44 \times 10^{-6}$ & $5.64 \times 10^{-4}$ \\
$10000.0$ & $135$ & $70$ & $0.9620$ & $0.0596$ & $1.18 \times 10^{-13}$ & $4.19 \times 10^{-6}$ & $5.51 \times 10^{-4}$ \\
%\hline
\bottomrule
\end{tabular}

\end{table}
%%%%%%%%%%%%%%%%%%%%%%%%%%%%%%%%%%

%%%%%%%%%%%%%%%%%%%%%%%%%%%%%%%%%%%%%%%%%%
%%%%%%%%%%%%%%%%%%%%%%%%%%%%%%%%%%%%%%%%%%
%%%%%%%%%%%%%%%%%%%%%%%%%%%%%%%%%%%%%%%%%%
%%%%%%%%%%%%%%%%%%%%%%%%%%%%%%%%%%%%%%%%%%
%%%%%%%%%%%%%%%%%%%%%%%%%%%%%%%%%%%%%%%%%%
%%%%%%%%%%%%%%%%%%%%%%%%%%%%%%%%%%%%%%%%%%
%%%%%%%%%%%%%%%%%%%%%%%%%%%%%%%%%%%%%%%%%%
%%%%%%%%%%%%%%%%%%%%%%%%%%%%%%%%%%%%%%%%%%
%%%%%%%%%%%%%%%%%%%%%%%%%%%%%%%%%%%%%%%%%%
\section{Conclusion}\label{conclusion}
The scenario of inflation was investigated in $f(R,T) = R + \eta T$ gravity theory. The theory, in a general view, could provide a non-minimal coupling between the matter field and the curvature by containing mixing term such as $R T$ in the action. Assuming a scalar field model, the matter sector of the action would be a combination of the scalar field and the kinetic term $\dot{\phi}^2$. Then, it could be assumed that any scalar field in this gravity theory is categorized as a new subclass of the k-essence model. Since the case of quintessence scalar field in the $R + \eta T$ gravity has been utilized for studying inflation, it sounds interesting and necessary step to modify the model and consider the inflation in such a theory for other scalar field models. In this manuscript, we study inflation in the $R + \eta T$ gravity theory by assuming the tachyon field as the inflaton. \\
The energy density and the pressure of the tachyon field are substituted in the dynamical equations and they were simplified by applying the slow-roll approximations. We then explored the scenario further by introducing three different types of potentials: power-law, generalized T-mode, and inverse hyperbolic cosine. Using Python coding and observational data, we determined the range of values for the free parameters of the model that allowed it to fit the data. \\

For the power-law potential, we found the two sets of values for the two parameters $n$ and $N$ for which to preserve the agreement with the data. It specified that the power value $n$ is required to be less than $1.3$ and the maximum of the number of e-fold is about $N = 47$. After plotting the density plot of the parameters, we found the values of $n$ and $N$ with higher probability density, implying $n = 0.25-0.70$ and $N = 32-37$. Considering the scalar power spectrum implies that the commoving modes we observe are related to the number of e-fold $N = 55 - 70$. However, this value is out of the acceptable set that was obtained for $N$. Therefore, there are no desirable consistency of the predicted scalar spectral index and the tensor-to-scalar ratio with the data. \\
Then, we obtain the sets of acceptable values for the parameters $\beta$ and $N$ for the case of generalized T-mode potential. By density plotting of the sets, we obtained the preferred values with higher probability, namely $\beta = 75-95 $ and $N = 70-85$. However, there was a good mount of data supporting the number of e-folds between $N = 65-70$ which indicates that there could be a good consistency between the model and the data and the model could be counted as a valid candidate for inflation. Considering the power spectrum implies that the values of the number of e-folds are related to the modes that would re-enter the Hubble horizon in the future time. The numerical result of the case was summarized in Table.\ref{GTmode_table}, where one could find the numerical prediction of the case about the scalar spectral index and the tensor-to-scalar ratio for different values of the model's parameters selected from the obtained region in Fig.\ref{GTmode_rns}. It is realized also the estimated energy scale of inflation is of the order of $\mathcal{O}(10^{-3}) M_p$.   \\
Finally, for the inverse hyperbolic cosh potential, we determined the plausible range for the parameters $\beta$ and the number of e-folds $N$. Among these values, the ones with higher probability are $\beta = 125-140$ and $N = 72-78$. Although the predicted range for the number of e-folds is our of our range of interest, there are still some data supporting the range $N = 65-70$ which brings the model to our attention. Since the power spectrum of the commoving modes that we are observing are related to ones that exited the Hubble horizon at the number of e-folds $N = 55-70$, this model with the inverse-cosh potential could be counted as a model of inflation. The estimated energy scale of inflation for the obtained free parameters was of the order of $\mathcal{O}(10^{-3}) M_p$, similar to the previous cases. The advantage of this case compared to the others was that the model was able to fit the data for a proper value of the number of e-folds. \\  

Within the frame of $f(R,T)$ gravity, the scenario of inflation has been considered in some prior works, where they only considered the canonical scalar field as inflaton. In \cite{Gamonal:2020itt}, where the scenario of inflation with a canonical scalar field as inflaton in $f(R,T) = R + 2 \kappa \lambda T $ is considered, the authors determined that the scalar spectral index and the tensor-to-scalar ratio remain unchanged and receive no change from the $f(R,T)$ gravity. Their result showed that the model does not have good consistency with the data. Here, we have the same situation with the tachyon scalar field when there is a power-law potential. Also, one could find the same situation in \cite{Ashmita:2022swc}, for a specific choice of the parameter. However, this result differs from \cite{Bhattacharjee:2020jsf}, where it is stated that the scalar spectral index and the tensor-to-scalar ratio are modified due to the $f(R,T)$ gravity. The situation is different when a different potential is used. The case of natural and hilltop potentials have been considered in \cite{Gamonal:2020itt} and it is realized that for these types of potentials, the coupling constant $\lambda$ changes the slow-roll parameters more as a redefinition of the potential constant. Similar behavior occurred for our model when we studied generalized T-mode and inverse-hyperbolic potentials. The redefined parameter $\beta$ is a combination of the model constants plus the coupling constant $\lambda$. \\  
The situation gets interesting where the mixing term, like $R T$, is included in the action. In this case, there will be a non-minimal coupling between the matter field and curvature. Such theory is investigated in \cite{Chen:2022dyq} where $f(R,T) = (1 + \alpha + \kappa^2 \beta \; T) R + \kappa \gamma \; T$. A canonical scalar field with chaotic and natural potentials has been considered, and it is found that the results are sensitive to the non-minimal coupling term, and there could be better compatibility with data compared to the gravity theory with no mixing term. Therefore, it is expected to get better results for tachyon inflation with power-law potentials when mixing terms are included.    \\

Overall, this study provides new insights into the scenario of tachyon inflation within the $f(R,T)$ gravity theory, highlighting its potential as a useful alternative framework.

%Gamonal:2020itt,Chen:2022dyq,Ashmita:2022swc,

%%%%%%%%%%%%%%%%%%%%%%%%%%%%%%%%%%%%%%%%%%
%%%%%%%%%%%%%%%%%%%%%%%%%%%%%%%%%%%%%%%%%%
%%%%%%%%%%%%%%%%%%%%%%%%%%%%%%%%%%%%%%%%%%
\section*{Acknowledgments}
The work of A.M. has been supported financially by “Vice Chancellorship of Research and Technology, University of Kurdistan” under research Project No.01/9/17886.

%%%%%%%%%%%%%%%%%%%%%%%%%%%%%%%%%%%%%%%%%%
%%%%%%%%%%%%%%%%%%%%%%%%%%%%%%%%%%%%%%%%%%
%%%%%%%%%%%%%%%%%%%%%%%%%%%%%%%%%%%%%%%%%%
\bibliography{REFtest}

%\begin{thebibliography}
%\bibitem{}
%\end{thebibliography}

\end{document}